\pdfoutput=1

\documentclass[final,fleqn,12pt,a4paper]{article}
\usepackage{a4wide,amsthm,amsfonts,amsopn,amsmath,amssymb,epsfig,graphics,verbatim}
\usepackage{dsfont}

\RequirePackage[pagebackref=true]{hyperref}
\hypersetup{colorlinks=true,linkcolor=blue,urlcolor=blue,citecolor=red,
    pdftitle=Rebuttal of On Nonparametric Identification of Treatment Effects in Duration Models by Johansson and Lee,
    pdfauthor=Jaap Abbring and Gerard van den Berg,
    pdfsubject=JEL Codes C14 C31 C41,
    pdfkeywords=econometrics survival analysis hazard rates scientific conduct 
    pdfdisplaydoctitle=true}

\title{Rebuttal of ``On Nonparametric Identification of Treatment Effects in Duration Models''\thanks{We are grateful to Arne Uhlendorff, Johan Vikstr\"om, Tiemen Woutersen and Olof \AA slund for interesting discussions.}} 

\author{Jaap H. Abbring\thanks{CentER, Department of Econometrics \& OR, Tilburg University, P.O. Box 90153, 5000 LE Tilburg, The Netherlands; and CEPR. E-mail: \href{mailto:jaap@abbring.org}{jaap@abbring.org}. Web: \href{http://jaap.abbring.org}{jaap.abbring.org}.}\and
Gerard J. van den Berg\thanks{University of Bristol, IFAU-Uppsala, IZA, ZEW, University of Mannheim, and CEPR.} 
}

\date{September 2016\\
Updated July 2019\footnote{The main text of this paper is identical to our IZA Discussion Paper 10248 from September 2016.  We have added an appendix in which we discuss related files currently (July 20, 2019) posted on M.J. Lee's web site. We note that these files simply reiterate Johansson and Lee's earlier claims, which were shown to be false both in our IZA Discussion Paper 10248 and, at least twice, in peer review at a major journal.
}}

\begin{document}
\maketitle

\begin{abstract}
\thispagestyle{empty}
\noindent In their IZA Discussion Paper 10247, Johansson and Lee claim that the main result (Proposition 3) in Abbring and Van den Berg (2003b) does not hold. We show that their claim is incorrect. At a certain point within their line of reasoning, they make a rather basic error while transforming one random variable into another random variable, and this leads them to draw incorrect conclusions. As a result, their paper can be discarded.

\end{abstract}

\newpage

\section{Introduction}

In a 2003 article in Econometrica (Abbring and Van den Berg, 2003b) we analyzed the specification and identification of causal multivariate duration models. We focused on the case in which one duration captures the point in time at which a treatment is initiated and one is interested in the effect of this treatment on some outcome duration. We defined a ``no anticipation of treatment'' assumption and showed that this assumption in a model framework inspired by the mixed proportional hazards model delivers identification. The model framework allows for dependent unobserved heterogeneity and does not require exclusion restrictions on covariates. In a nutshell, the timing of events conveys useful information on the treatment effect.

In their IZA Discussion Paper 10247, Johansson and Lee (henceforth JL) claim that our main identification result (Proposition 3, for our Model 1A) does not hold.\footnote{\label{fn:proposition}JL do not specifically refer to our Proposition 3. However, halfway their page 4, they do refer to the ``main identification finding of [Abbring and Van den Berg (2003b)] (pp. 1505--1506)'', and our Proposition 3 is the only relevant formal identification result on the pages 1505--1506 of the Econometrica article. Moreover, on page 13 of JL, they refer to our Model 1A, and Proposition 3 is the main identification result for Model 1A.} JL examine two model versions, which they refer to as DGP1 and DGP2. It is good custom that comments on articles from other researchers phrase their statements in terms of the notation and model specification of the original article. However, JL do not make this effort. They claim that their DGP1 captures our model (in particular, our Model 1A) and that our proof of Proposition 3 does not apply to DGP1.  From this, they conclude that our proof is incorrect and Proposition 3 is not true.

In this note we show that JL's claims are incorrect. They make a rather basic error when deriving their DGP1 from our Model 1A. Consequently, DGP1 and our Model 1A are not equivalent, and the fact that our proof of Proposition 3 does not apply to DGP1 has no bearing on its validity. We show that, in fact, their DGP2, for which they do not raise similar concerns, is consistent with our model, in contrast to JL's claims.

In Section \ref{s:error} we develop the setup in terms of JL's notation. We discuss properties of our Model 1A and show that these immediately refute statements early on in JL about our identification analysis. In Section \ref{s:error} we explain the key error in JL's derivations. Section \ref{s:concl} concludes and critically assesses awkward remarks in JL about the empirical researchers who cited our 2003 Econometrica article.

\section{Johansson and Lee's setup and notation}
\label{s:setup}

The setup, in JL's notation, is a model with a treatment duration $W$ and an outcome duration $Y$ (throughout, following JL, we will keep observed covariates and unobserved heterogeneity implicit as these are not relevant to the argument). The treatment duration $W$ has a hazard rate $h_W(w)$ and Lebesgue density
\begin{equation}
\label{eq:fw}
f_W(w)=h_W(w)\exp\left(-\int_0^w h_W(\tau)d\tau\right)
\end{equation}
at duration $w$. The outcome duration $Y$ given $W=w$ has a hazard rate $h_0(y)$ at times $y\leq w$ and a hazard rate $h_1(y)$ at times $y>w$, with corresponding Lebesgue density
\begin{equation}
\label{eq:fy}
f_{Y|W}(y|w)=\left\{\begin{array}{lll}
    h_0(y)\exp\left(-\int_0^y h_0(\tau)d\tau\right)&\text{if}&y\leq w\text{    and}\\
    h_1(y)\exp\left(-\int_0^w h_0(\tau)d\tau-\int_w^y h_1(\tau)d\tau\right)&\text{if}&y>w.\\
\end{array}\right.
\end{equation}
The joint density of $(W,Y)$ at $(w,y)$ is $f_{Y|W}(y|w)f_W(w)$.

In our 2003 Econometrica article, we were specifically interested in the differences between the hazard rates $h_0$ and $h_1$, which capture the ``treatment effects'' of the event at time $W$ on the outcome duration $Y$. Therefore, we provided results on the identification of $h_0$ and $h_1$ under general conditions. To this end, we used that our Model 1A embeds an independent competing risks model with hazard rates $h_W$ and $h_0$ for the so-called {\em ``identified minimum''} of $W$ and $Y$ (i.e., their minimum {\em and} whether this minimum is equal to $W$ or equal to $Y$). That is, we used that the subdensities of $Y$ on $\{Y<W\}$ and $W$ on $\{W<Y\}$ (which, together, fully characterize the distribution of this identified minimum) equal
 \begin{equation}
\label{eq:y}
-\frac{d}{d y}\Pr\left(Y>y,Y<W\right)=h_0(y)\exp\left(-\int_0^y h_0(\tau)d\tau-\int_0^y h_W(\tau)d\tau\right)
\end{equation}
and 
 \begin{equation}
\label{eq:w}-\frac{d}{d w}\Pr\left(W>w,W<Y\right)=h_W(w)\exp\left(-\int_0^w h_W(\tau)d\tau-\int_0^w h_0(\tau)d\tau\right),
\end{equation}
respectively (see also JL's (2.4)(i) and (2.4)(ii)). This intermediate result is key because it allowed us to apply Abbring and Van den Berg's (2003a) identification result for the competing risks model to establish the identification of $h_W$ and $h_0$ (as well as, in the general setting with observed covariates and unobserved heterogeneity, the identification of the covariate effects on these hazard rates and the distribution of the unobserved heterogeneity in them) from data on the identified minimum of $W$ and $Y$. Abbring and Van den Berg (2003a) proved a one-to-one mapping between the competing risks model and the distribution of the identified minimum. As is intuitively clear, and apparent from Equations (\ref{eq:y}) and (\ref{eq:w}), the distribution of the identified minimum of $W$ and $Y$ only involves treatment and outcome hazards before either the treatment or the outcome event occurs and does not depend on the post-treatment outcome hazard (or, for that matter, the post-outcome treatment hazard). Indeed, the specification of the hazard rates after the minimum of $W$ and $Y$ is irrelevant for the competing risks identification result. The competing risks model in Abbring and Van den Berg (2003a) and the competing risks part of our Model 1A are equivalent.\footnote{Underlying this equivalence is the no-anticipation assumption for potential outcomes made in Abbring and Van den Berg (2003b), which ensures that the potential outcome hazards before treatment do not depend on the eventual treatment time.} In particular, in Model 1A (without covariates and unobserved heterogeneity), the implied submodel for the identified minimum of $W$ and $Y$ is an {\em independent} competing risks model, despite the fact that $W$ and $Y$ will generally be dependent if $h_0$ and $h_1$ differ.

JL state halfway their page 4 that $h_0=h_1$ (``no treatment effects'') is {\em necessary} for our key intermediate result that Model 1A embeds an independent competing risks model with hazard rates $h_W$ and $h_0$. This is false. This should be clear from the previous paragraphs, but we can illuminate it further by examining the subsurvival functions $\Pr(Y>y,Y<W)$ of $Y$ on $\{Y<W\}$ and $\Pr(W>w,W<Y)$ of $W$ on $\{W<Y\}$, in order to verify that Equations (\ref{eq:y}) and (\ref{eq:w}) hold for general $h_0$ and $h_1$, including cases where $h_0\neq h_1$. The subsurvival function of $Y$ on $\{Y<W\}$ satisfies
 \begin{equation}
 \label{eq:subsurv}
\begin{split}
\Pr&\left(Y>y,Y<W\right)\\
&=\int\Pr\left(Y>y,Y<W|W=w\right)f_W(w)dw\\
                &=\int_y^\infty\Pr\left(y<Y<w|W=w\right)f_W(w)dw\\
                &=\int_y^\infty \int_y^w f_{Y|W}(\tau|w)d\tau f_W(w)dw\\
                &=\int_y^\infty \left[\exp\left(-\int_0^y h_0(\tau)d\tau\right)-\exp\left(-\int_0^w h_0(\tau)d\tau\right)\right]f_W(w)dw.
\end{split}
\end{equation}
Differentiating the left-hand and right-hand sides of (\ref{eq:subsurv}) with respect to $y$, substituting (\ref{eq:fw}), integrating, and multiplying by $-1$ gives (\ref{eq:y}). An analogous derivation for the subsurvival function of $W$ on $\{W<Y\}$ gives (\ref{eq:w}). Consequently, (\ref{eq:y}) and (\ref{eq:w}) hold generally, and this confirms that, in contrast to what JL claim halfway their page 4,  Abbring and Van den Berg's (2003a) results for the competing risks model can be applied to the model for the identified minimum of $W$ and $Y$ embedded in Abbring and Van den Berg's (2003b) model of treatment effects. 

\section{Johansson and Lee's error}
\label{s:error}

It is easy to see where JL go wrong in their own derivations. Halfway their page 6, they claim that we have adopted DGP1, which they associate with their Equation (2.2). For a proof, they refer to their Appendix. In this Appendix, at the top of page 13, they specify the hazard rate of the potential outcome $Y^*(w)$, evaluated at the elapsed duration $y$, as $h_0(y)$ if $y\leq w$ and as $h_1(y)$ if $y>w$. This hazard rate specification indeed corresponds to our Model 1A. The subsequent equation in their Appendix is supposed to capture $Y^*(w)$ as the inverse of the corresponding integrated hazard evaluated in a standard exponential random variable, as is clear from their reference to the equation halfway Abbring and Van den Berg (2003b, p. 1496) and the corresponding discussion below their Equation (2.2). However, the integrated hazard that is used here is not the correct integrated hazard, because it is not the integral of the hazard in the equation at the top of page 13. Specifically, if $w>Y^*(w)$ then the integrated hazard should equal $\int_0^{Y^*(w)} h_0(\tau)d\tau$. Instead, in this case, JL take it to equal $ \int_0^w h_0(\tau)d\tau - \int_{Y^*(w)}^w h_1(\tau) d\tau$.\footnote{From JL's calculations for the special case with constant hazards, in particular their derivation of (3.6) in their Appendix, it is clear that that they indeed interpret $\int_w^{Y^*(w)}h_1(\tau) d\tau$ as $-\int_{Y^*(w)}^w h_1(\tau) d\tau$  if $w>Y^*(w)$.} Of course, this leads to a very peculiar DGP1 with absurd implications. However, DGP1 is not consistent with our Model 1A and its absurd implications are solely due to the mistake by JL and do not carry over to our Model 1A.

If JL would have used the correct integrated hazard instead, they would have ended up with their DGP2 instead of DGP1. To see this, note that their Equation (2.6)(i) applies if 
\[
\exp\left[-\int_0^Wh_0(\tau)d\tau-\int_W^Yh_1(\tau)d\tau\right]\leq \exp\left[-\int_0^Wh_0(\tau)d\tau\right]\Longleftrightarrow Y\geq W
\]
and that their Equation (2.6)(ii) applies if 
\[
\exp\left[-\int_0^Y h_0(\tau)d\tau\right]>\exp\left[-\int_0^Wh_0(\tau)d\tau\right]\Longleftrightarrow Y<W,
\]
where we have used JL's assumption that $h_0(\tau)>0$  and $h_1(\tau)>0$ for all $\tau$ (page 4). Moreover, the left-hand side of Equation (2.6)(i) involves the correct integrated hazard of $Y$ evaluated at random $W$ and $Y$ such that $Y\geq W$, $\int_0^Wh_0(\tau)d\tau+\int_W^Yh_1(\tau)d\tau$,  and the left-hand side of Equation (2.6)(ii) involves the correct integrated hazard of $Y$ evaluated at random $W$ and $Y$ such that $Y\leq W$, $\int_0^Y h_0(\tau)d\tau$ (note that both are identical and correct in the boundary case $Y=W$). Consequently, in contrast to what JL claim, their DGP2 specified by Equation (2.6) is consistent with our Model 1A. Moreover, as JL note halfway page 6, under DGP2, (\ref{eq:y}) and (\ref{eq:w}) hold for general $h_0$ and $h_1$. Taken together, this implies that JL's key claim that (\ref{eq:y}) and (\ref{eq:w}) cannot be used in the identification analysis of Model 1A is wrong.

Halfway page 6, JL raise a new concern about their DGP2, and thus our Model 1A: Information on the post-treatment outcome hazard $h_1$ can only be obtained from the selected subpopulation with $Y>W$. We share this concern; in fact, this is one aspect of the selection problem that is at the core of our paper and that we addressed successfully in it.\footnote{Other, more subtle aspects of this problem concern the selection on the unobservable heterogeneity factors that is kept implicit in JL and here.} JL do not show that this concern invalidates our identification results, in particular our Proposition 3. Rather, they propose to infer some treatment effect parameters by regressing $Y$ on $W$ in a subsample of the population with $Y>W$ and claim that the resulting estimator is inconsistent. We never proposed this ad hoc procedure and we would certainly not recommend it. In any case, its failure to produce a consistent estimator of certain treatment effects does not prove our identification results wrong.

In sum, JL make a basic error in deriving their DGP1 from our Model 1A. This leads them to incorrectly claim that their DGP1 and our Model 1A are equivalent. Instead, our Model 1A corresponds their DGP2, to which their concerns about our identification analysis do not apply.  After reading JL, one may wonder why these misunderstandings have not surfaced in an earlier stage of the publication process. Indeed, when presenting their claim that we adopted DGP1 in Abbring and Van den Berg (2003b), JL note halfway their page 14 that ``[Abbring and Van den Berg] did not object to DGP1, nor did they suggest DGP2''. Now, JL have not offered us the opportunity to respond to the most recent version of their note before they submitted it for publication as an IZA Discussion Paper. We did communicate with JL about previous drafts of their paper in 2014, both directly, after they sent us their paper's first draft in January 2014, and indirectly through the editorial process at a journal. In these communications, we pointed out a logical flaw in an argument they used at the time, and that flaw has disappeared from the current draft (they claimed that $h_0=h_1$ is necessary for (\ref{eq:y}) and (\ref{eq:w}) to hold but only showed it to be sufficient). We also directly demonstrated that their claim that $h_0=h_1$ is necessary for (\ref{eq:y}) and (\ref{eq:w}) to hold is incorrect, by providing, as in this Rebuttal, elementary calculations that show that these equations hold generally. At the time, we did not provide an exhaustive list of all aspects of their paper that we disagreed with, because this was not necessary to make our point that they were wrong. In particular, we did not specifically refer to JL's DGP1 in our earlier private communication with them. The reader can rest assured, though, that both of us disagree with many things that we never explicitly mentioned or objected to.

\section{Conclusion}
\label{s:concl}

The claims in JL's paper about the results in Abbring and Van den Berg (2003b) can be discarded.

In their paper, JL observe that our paper has been often cited by empirical studies, and they mention a number of authors who cited our work. In the light of what JL perceive as an incorrectness in our paper, they speculate about the reason for the high citation score. Specifically, they claim that ``the most likely reason is that Abbring and Van den Berg (2003b) is a difficult paper to read and the applied studies in the literature simply took the finding at the face value''. We view this as a preposterous statement. It offends the empirical researchers who cite our work, by depicting them as simple minds unable to understand methodological work. Our current note provides a better reason: our 2003 finding is correct.

\newpage

\section*{References}
\newlength{\leftlocal}
\setlength{\leftlocal}{\leftmargini} \addtolength{\leftmargini}{-.5\leftmargini}

\begin{description}
\newlength{\labellocal}
\setlength{\labellocal}{\labelwidth} \setlength{\labelwidth}{5pt}
\newlength{\itemlocal}
\setlength{\itemlocal}{\itemsep} \setlength{\itemsep}{0pt} 

\item Abbring, J.H. and G.J. van den Berg (2003a), ``The identifiability of the mixed proportional hazards competing risks model'', \textit{Journal of the Royal Statistical Society Series B}, 65,  701--710.
  
\item Abbring, J.H. and G.J. van den Berg (2003b), ``The non-parametric identification of treatment effects in duration models'', \textit{Econometrica}, 71, 1491-1517.

\end{description}

\clearpage
\appendix

\section*{Appendix. Update on files currently on Lee's web site}

At \href{https://sites.google.com/site/mjleeku/working-papers}{sites.google.com/site/mjleeku/working-papers}, M.J. Lee currently (July 20, 2019) posts an old (March 11, 2015) version of the September 2016 IZA Discussion Paper 10247 that this Rebuttal shows to be incorrect. The only difference between the two papers seems to be that the last paragraph on page 10 in the March 2015 paper does not appear in the September 2016 IZA Discussion Paper. Our Rebuttal does not refer to this paragraph, so applies without change to the March 2015 paper on Lee's website (except that the hazard rate ``at the top of page 13'' is now halfway page 13, etcetera).

Lee's web site also offers a more recent ``Re-rebuttal'' (December 2, 2016), in which JL respond to our Rebuttal. The main argument in this Re-rebuttal, on its page 3, is essentially the same as the argument in the first part of their paper's Appendix. In Section \ref{s:error}, we have already explained in detail why this argument is incorrect. To be sure, we reiterate our main objection to JL's argument here.

JL claim that our main identification result (Proposition 3, for our Model 1A) does not hold (they are not explicit about this, but see our Footnote \ref{fn:proposition}). JL seem to agree with us that a homogeneous special case of our Model 1A (in JL's notation) specifies the hazard rate of the ``treatment'' duration $W$ at time $w$ as $h_w(w)$ and the hazard rate of the ``outcome'' duration $Y$ given treatment at time $W$ as $h_0(y)$ at times $y\leq W$ (before treatment) and as $h_1(y)$ at times $y>W$ (after treatment). In our original identification analysis of Model 1A, we directly derived this specification's implications for the joint distribution of $(W,Y)$ using that the integrated hazard of $Y$ at $y$ given $W$ equals $\int_0^y h_0(\tau)d\tau$ if $y\leq W$ and $\int_0^W h_0(\tau)d\tau+\int_W^yh_1(\tau)d\tau$ if $y>W$ (which follows directly from the homogeneous special case of our model). As we pointed out in Section \ref{s:error}, JL instead use $\int_0^W h_0(\tau)d\tau-\int_y^W h_1(\tau)d\tau$ for the integrated hazard of $Y$ given $W$ at $y<W$. This does not follow from our model, makes no sense, and is the reason JL cannot reproduce our results. 

Note that JL use a more indirect approach to study our model's implications for the distribution of $(W,Y)$ than we did. They specify the ``DGP'' corresponding to our Model 1A (we did not refer to a ``DGP'') by constructing $W$ and $Y$ from independent uniform random variables using the inverse probability transform (we used a similar construction earlier in the paper to relate our analysis to the potential-outcomes framework in the  treatment-effects literature, but not in our analysis of Model 1A). Eventually, this should however lead to the same conclusions. It doesn't, because they use an incorrect distribution function, based on an incorrect integrated hazard, in their construction.

In sum, JL's arguments continue to be fatally flawed. Indeed, like a 2014 version of JL's paper, the paper and Re-rebuttal currently on Lee's web site were peer reviewed at a major economics journal and rejected for being incorrect. 
\end{document}